# Dispersion and Thickness Control in Evaporation-induced Self-Assembly of Opal Photonic Crystals

Alex Grant[1], Alex Lonergan[1], and Colm O'Dwyer[1,2]*

[1]School of Chemistry, and Environmental Research Institute, and Tyndall National Institute, University College Cork, Cork, T12 YN60, Ireland

[2]Advanced Materials and BioEngineering Research Centre, Trinity College Dublin, Dublin 2, Ireland

**Abstract**

Opals are naturally occurring photonic crystals which can be formed easily using low-cost self-assembly methods. While the optical behaviour of opals has received significant attention over the last number of decades, there is limited information on the effect of crystal thickness on the optical properties they display. Here, the relationship between volume fraction and crystal thickness is established with an evaporation-induced self-assembly (EISA) method of formation. The extent to which thickness can be used to manipulate the optical properties of the crystals is explored, focusing on the change in the photonic band gap (PBG). Microscopical structural characterization and angle-resolved transmission spectroscopy are used to examine the quality of the photonic crystals formed using different volume fractions of polystyrene spheres, with thicknesses up to 37 layers grown from volume fractions of 0.125%. This work provides a direct correlation between sphere solution volume fraction and crystal thickness, and the associated optical fingerprint of opal photonic crystals. Maximum thickness is examined, which is shown to converge to a narrow range over several evaporation rates. We identify the criteria required to achieve thickness control in relatively fast evaporation induced self-assembly while maintaining structural quality, and the change to the spectroscopic signature to the (111) stopband and higher order (220) reflections, under conditions where a less ordered photonic crystals are formed.

**Introduction**

For researchers interested in perfecting colloidal crystal growth, maximizing structural quality requires optimization of an array of environmental conditions.(1-6) Bottom-up self-assembly methods such as dip coating,(7) spin coating,(8) Langmuir Blodgett deposition,(9) electromagnetic-induced self-assembly,(10) and evaporation-induced self-assembly (EISA) are simple and cost effective, but difficult to control due to their spontaneous nature. As has been demonstrated, colloidal crystal films involving silica or polymeric spheres as opal, suffer from localized regions of disorder, layer-on-layer registry changes that affect long range order, and cracking from various forces during drying. There have been attempts to control these issues recently.(11-13)

There have been approaches that allow 2D colloidal crystals to form from a single layer of spheres, showing diffraction grating reflection characteristics.(14) More recently, 2D colloidal crystal and opals were shown to be transferred to more useful (and often complex or problematic substrates) by capillary transfer to give very high quality coatings useful for application.(15) A key advantage of many of these approaches is fast growth of potentially ultra-thick colloidal crystals, or the formation of functional films with required or useful optical properties, on substrates that are best for the application. The requirements for high crystal quality are now well established. The conditions which must be controlled to affect crystal growth are also documented. However, some key questions still remain. Under what conditions and to what extent can crystal thickness be directed while maintaining overall structural quality?

Knowing the factors that influence colloidal crystal growth of opal photonic crystals from relatively simple solution-based evaporation methods is important.(16) These amenable photonic crystals are well known to manipulate photons, like the manipulation of electrons by the atomic lattice.(17-19) Consisting of periodic dielectric material, the structure can be designed to inhibit light propagation in specific directions, providing the ability to tune the frequencies which are allowed to propagate, with a range of potential applications in fields such as optics and optoelectronics,(20) energy storage,(21, 22) and sensors for environmental monitoring, biomedicine and food preservation.(23) A photonic crystal can also be designed such that light propagation is inhibited in all directions at specific frequencies, and crystal or film quality is important here, to minimize scattering and dispersion losses, for example. The omnidirectional photonic band gap (PBG) of a 3D photonic crystal is derived from Maxwell's equations.(24)

Perfectly ordered self-assembled opals can be constructed entirely by repetition of the face-centered cubic (FCC) unit cell, the most thermodynamically favourable packing arrangement of crystalline spheres.(25) For an ideal opal, the FCC unit cell is repeated with the spheres stacked in the direction normal to the lattice plane described by the miller indices (111).(27, 28) The wavelength-dependent structural colour observed in opals can be approximated well by the Bragg-Snell law that results from an extension of the dispersion relations.(30)

Despite a tendency towards (111) FCC stacking during opal self-assembly, defective areas are inevitable. Stacking faults are most common, and usually occur in the (100) crystallographic direction. Surface SEM images can be used to distinguish between the two stacking arrangements. Planes corresponding to the (111) and (100) directions are observed as hexagonal and square patterns, respectively. A host of other stacking arrangements are possible, but more difficult to detect.(32) In optical transmission spectra, (111) stacking arrangements are observed as the primary transmission dip constituting the PBG of the opal. Stacking faults such as in the (100) direction occur as transmission dips with a significantly lower depth relative to the PBG towards the higher energy region of the spectrum.

Forms of disorder that significantly affect the optical spectra of opals include grain boundaries, defects, and dislocations of unit cells and polydispersity. These processes lead to a range of scattering regimes, each of which impacts the intensity profile of the optical spectra depending on the frequency range under consideration.(33) While there is very often a statistically quantifiable average variance in sphere diameter, the standard deviation must be minimal to prevent destruction of the spectral shape of the PBG in a high-quality colloidal opal film as it has been shown that polydispersity and displacements of unit cells determine scattering loss in 3D photonic crystals.(34) Such deviations from perfect periodicity cause exponential attenuation of the coherent beams, the propagation of which generate the PBG. The attenuation affects the mean free path, also known as the extinction length. After propagation over this length the coherent light beams are converted to a diffuse glow which impacts the associated optical spectra and prevents the application of such materials in any photonic integrated circuit.(35) It is instructive to know how the factors of the growth such as colloid concentration, bath temperature and thus the crystallization or evaporation rate, and final thickness, affect degree of order.

We use (EISA) to study the spectroscopic signature of the PBG in opals films. Both order and thickness can be controlled primarily by the sphere solution volume fraction and evaporation rate. Previous research indicates that faster evaporation rates can result in earlier onset crystallization, which causes a more disordered surface structure.(37) On the other hand, higher kinetic energy provided at these faster rates mean that the spheres can explore more favourable lattice sites, leading to a greater overall tendency towards FCC packing. Maximal crystal quality and thickness control is achieved through the selection of the optimal evaporation rate and volume fraction combination.

Here, we establish the relationship between volume fraction and opal thickness at three evaporation rates formed using EISA. In tandem with electron microscopy, we characterize the degree of order in successively thicker films up to 37 layers by examining the modification to dispersion using angle-resolved transmission spectra with Bragg-Snell analysis. Lastly, the assembly conditions under which the crystals can no longer be characterised using traditional Bragg-Snell analysis applied to the PBG are identified.

**Experimental**

**Materials and Substrate Preparation**

Colloidal solutions of polystyrene (PS) spheres in 2.5 % (w/v) aqueous suspension with diameters of 370 nm and 500 nm were purchased from Polysciences Inc. The sphere suspension possessed a net negative charge due to functionalization using a sulfate ester during formation to aid in self-assembly. The substrates used were fluorine-doped tin oxide (FTO)-coated soda-lime glass purchased from Solaronix SA. Glass of thickness 2.2 mm was cut into pieces with dimensions 10 × 25 mm$^2$. The conductive layer of the pieces consisted of a thin FTO layer followed by a thin $SiO_2$ layer (each layer ~20 nm) followed by a thicker FTO layer (~300 nm). The conductive layer rested upon a glass substrate. All substrates were cleaned using the same process: Successive sonication in acetone (reagent grade 99.5%; Sigma Aldrich), isopropyl alcohol (reagent grade 99.5%; Sigma Aldrich) and de-ionized water, each for 5 minutes. All cleaned samples were dried at room temperature in an inert atmosphere. Kapton tape was used to cover the back of the substrate and partially cover the surface of the substrate such that an exposed surface of 10 × 10 mm$^2$ remained. This ensured colloidal crystal formation occurred only on the desired surface area. The samples were treated in a Novascan PSD Pro Series digital UV-ozone system for 1 h to maximize hydrophilicity of the surface of the substrates immediately prior to photonic crystal formation.

**Evaporation-Induced Self-Assembly**

Opals were formed using evaporation-induced self-assembly. FTO glass substrates were immersed in a diluted PS sphere solution with volume fractions in the range of 0.025% to 0.125 % which were placed on a hot plate at a constant temperature between 28 °C and 34 °C. Once the sphere solution was evaporated fully, the substrates were removed and left to dry in inert atmosphere at room temperature. To ensure consistent EISA of the colloidal films at a well-defined angle, we used Vat polymerization 3D printing to create beakers that were formed from photopolymerized poly(methyl methacrylate)-based resin using a Formlabs Form 2 stereolithography printer. Further details are provided in the Supplementary Information, Fig. S1. All beakers were cleaned using the same sonication process described above, followed by UV-ozone treatment for 1 h. The diameter of each beaker was 6 cm and substrate rests were positioned at an angle of 30 ° to the beaker walls.

**Microscopy and Spectroscopy**

All microscopy was performed using a FEI Quanta 650 SEM at accelerating voltages of 10 kV to 30 kV. Prior to imaging, an incision was made along the length of each sample using a thin blade so that the cross-section could be imaged for thickness measurements and were then sputtered using an ATC Orion−5−UHV sputtering system to increase conductivity and improve grounding of the sample. Gold coating on non-conductive PS spheres was carried out using a Quorum 150T S magnetron sputtering system. Sputtering samples reduced charging effects, namely electron diffusion and beam drift,(38) allowing better contrast and feature definition. For a thickness line measurement $L$, the actual thickness is calculated according to $T = L/\cos(45°) = L\sqrt{2}$. Thickness measurements were enabled by tilting the stage on which the sample lay parallel to the surface at an angle of $45°$. For each sample, several thickness measurements were obtained and the mean values in each case used as data points to establish the volume fraction-thickness relationship.

Angle-resolved transmission spectra were used to characterise colloidal opal films. The light source was a tungsten-halogen lamp with an operating wavelength range of 400−2200 nm. Transmission measurements were recorded using a UV-visible spectrometer (USB2000 + VIS-NIR-ES) with an operational range of 350−1000 nm and a NIR spectrometer (NIRQuest512-2.5) with an operational range of 900 nm to 2500 nm, both from Ocean Optics Inc. Adjustments to the incident angle for transmission measurements was enabled by a motorized rotational stage (ELL8; Thorlabs Inc.).

## Results and Discussion

Several colloidal opal films were grown using the EISA method in a range of sphere solutions and at several temperatures. The relationship between sphere solution volume fraction and opal thickness is shown in Fig. 1 (a) for the three temperatures investigated. The evaporation times were approximately 24 h, 36 h, and 48 h for temperatures of 34°C, 31 °C, and 28 °C, respectively. The cross-sectional SEM images used for all thickness measurements are shown in Fig. S2.

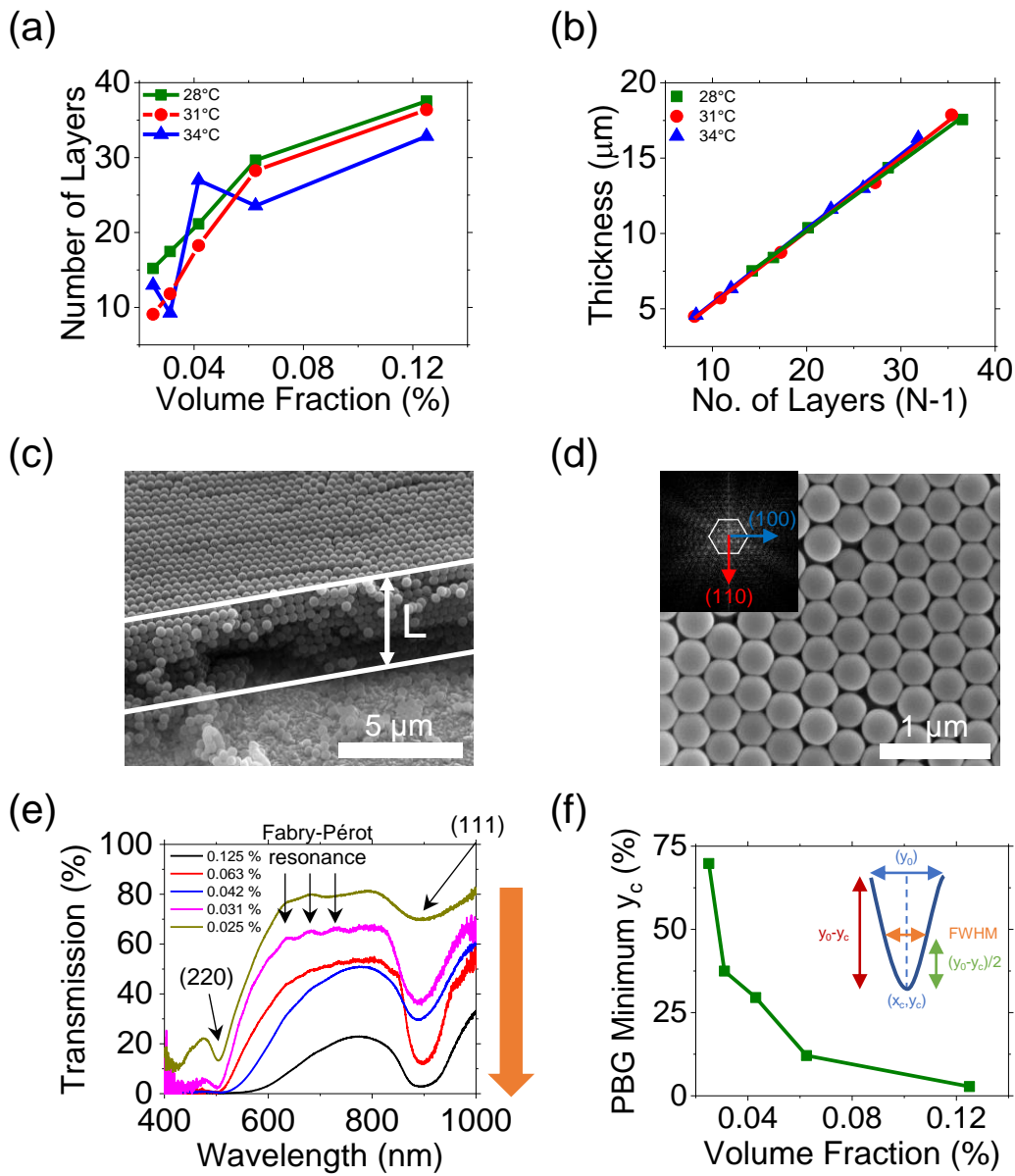

**Figure 1.** (a) Volume fraction-thickness relationship for opals grown from PS sphere solutions of five volume fractions at three evaporation temperatures. (b) Number of opal layers calculated using Eq. 3 as a function of measured sample thickness for samples formed under all investigated conditions. (c) Cross-sectional SEM image of representative opal demonstrating thickness measurement. (d) Plan view SEM image of surface of an opal grown from a volume fraction of 0.125 % at 28 °C. Inset: FFT with 2D hexagonal symmetry, the (100) and (110) directions labelled corresponding to the Γ-X and Γ-L directions of a 2D hexagonal lattice. (e) Transmission spectra obtained at normal incidence for opals grown from all five volume fractions at 28 °C. (f) Transmission intensity at the PBG centre as a function of volume fraction for all opals grown at 28 °C. Inset: Schematic illustrating the defining parameters of spectral transmission dips.

At all three temperatures, large crystal thicknesses for each volume fraction are achieved relative to those observed in the literature. In general, slower evaporation rates provide time for the self-assembly process and stacking of layers to continue at each point along the direction of evaporation flux. Excluding two outlier cases, the two lowest volume fractions grown at 34 °C, crystal thickness for each volume fraction at a given temperature is larger than the equivalent volume fraction at a higher temperature. Measured opal thicknesses range from a minimum of ~4.5 μm (9 layers) to a maximum of ~17.9 μm (37 layers). The investigated temperature range is narrow at only 6 °C, yet the evaporation time of the assembly process at the lowest temperature of 28 °C is twice that at the highest of 34 °C. A relatively small change in temperature significantly affects the speed of colloidal self-assembly for the system used here.

For the majority of EISA experiments found in the literature, small sphere solution surface areas are incorporated to minimize convective air flow and allow slow evaporation rates at higher temperatures of 60 – 80 °C.(39, 40) In our experiments evaporation occurs over relatively large solution surface areas. Given that molecules contained in a liquid evaporate at the gas-liquid interface,(41) evaporation rate is proportional to the exposed surface area of the solution during the assembly process. To provide context, in one study, the EISA process was performed at 65 °C over a period of 1-2 days. In our experiments, the EISA process was performed at 28-34 °C over 1-2 days. Crystal growth occurs at a similar rate at approximately half the temperature. The comparatively large surface area is helpful to both crystal quality and thickness. The result of our methodology is the assembly of multi-layered crystals of large thicknesses up to 37 layers for comparatively low volume fractions relative to the literature with a high level of order above a minimum evaporation rate and volume fraction combination.(42, 43)

While a general increase in thickness as a function of volume fraction is observed for each temperature, the increase is not linear (Fig. 1(a)). The height, $h$ of an opal sphere stack can alternatively be estimated using a simple geometric approach such as

$$h = D + (N-1)\sqrt{\frac{3}{2}}D \qquad (3)$$

where $D$ is the PS sphere diameter and $N$ is the number of opal layers. Fig. 1 (b) shows a scatter plot of measured thickness as a function of $N-1$, calculated using Eq. 3 for all films grown for each sphere solution volume fraction at each of the three evaporation temperatures. The lines correspond to the line of best fit for each dataset. Comparison of the parameters obtained from the line of best fit in tandem with Eq. 3 with measured experimental data determines the accuracy of the approach in predicting crystal thickness and sphere diameter. Based on Eq. 3, the *y*-axis intercept from each line of best fit can be approximated as the average opal sphere diameter *D*. The slope of the line approximates the interplanar spacing $d = \sqrt{\frac{3}{2}}D$ of the opals based on fcc stacking in the [111] direction, a stacking arrangement that is proven dominant based on the Bragg-Snell analysis of spectral data provided in Fig. S3.

Calculations of average sphere diameter *D* and average interplanar spacing *d* at each evaporation temperature using Eq. 3 were performed. The calculated and corresponding experimentally measured values are included in Table S1. For the interplanar spacing *d* at the higher two evaporation temperatures, there is clear agreement, with a relative difference of less than 1% between measured and values calculated from the slope of the lines of best fit. For the lowest evaporation temperature, there is a relative difference of 6 %. For the average sphere diameter *D*, the relative difference between experimental measurements and calculations using the *y* intercepts of the lines of best fit in order from highest to lowest evaporation temperature is 11%, 10%, and 163%. For the higher two temperatures the approximation is relatively accurate. For the lowest temperature, the approximation is completely inaccurate. The accuracy of the approximation of sphere diameter *D* using the y intercept of Eq. 3 can be further analysed by an examination of the practical considerations of each value. In Eq. 3, the quantity *N* corresponds to the number of opal layers. The *y* axis in Fig. 1 (b) corresponds to (*N* - 1). Therefore, at the *y* intercept where *N* - 1 = 0, *N* = 1 opal layer. Practically, 1 opal layer should be equal to the opal sphere diameter. For the two higher evaporation temperature datasets, the approximations are practically realistic to approximately 10% of the measured average sphere diameter. For the lowest temperature, applying Eq. 3 to the linear fit in Fig. 1 (b) is inaccurate and this approach is not applicable. The inaccuracy of the data corresponding to the highest evaporation temperature in Fig. 1 (b) can be attributed to the two outlier points corresponding to the two lowest volume fractions which indicates this evaporation rate is too fast to achieve thickness control below a volume fraction of 0.063%.

In Fig. 1 (c), a cross-sectional SEM image of a representative opal sample is shown, illustrating the approach used to obtain thickness measurements. A plan view SEM image of another representative opal is provided in Fig. 1 (d), demonstrating the order at the surface of the sample. The inset shows the corresponding FFT image, emphasising the hexagonal surface symmetry. The [100] and [110] directions are highlighted, given that opal layer growth occurs perpendicular to the (111) plane of an fcc lattice.

Optical transmission spectra for opals grown from all five volume fractions at an evaporation temperature of 28 °C and an incident angle range of 0–25 ° are shown in Fig. 1 (e). The primary transmission dip formed from light interaction with the (111) fcc lattice plane is observed in each case, corresponding to the PBG of each opal. As will be discussed later, opals grown at higher volume fraction show a stronger transmission suppression at ~500 nm corresponding to the (220) planes of the photonic crystal.(44)

While stacking faults along the (111) direction are the most common defects in artificial opals, the optical properties are unaffected at low energies such as those close to the PBG. Disorder in the form of grain boundaries, point defects, and dislocations affect the optical properties in two ways depending on whether the considered spectral region lies outside the corresponding stopband, the PBG in this case, or is contained within. Outside the PBG, light propagation becomes diffuse in a cone around the zero order beam.(45) Such diffuse scattering is observed here by the  in the form of Fabry-Perot resonances close to the blue edge of the PBG at the three lower volume fractions, but is suppressed for the two largest volume fractions. Inside the stopband region, there is an interplay between the coherent Bragg scattering which forms the stopband, and incoherent diffuse scattering leads to enhanced diffuse intensity for frequencies near the edges of the stopband, which is usually observed in optical spectra as an attenuation of the stopband. Examination of stopband attenuation for crystals of equivalent thickness can be used to indicate the magnitude of diffuse scattering.  Here, the optical spectra are presented for a range of different thicknesses, so other scattering regimes must be considered when interpreting the evolution of the optical spectra.

Figure 1 (f) shows the measured transmission intensity $y_c$ at the central frequency $x_c$ at normal incidence of the PBG for opals grown at an evaporation temperature of 28 °C as a function of sphere solution volume fraction. An exponential decay in $y_c$, and thereby thickness, increases. While order is sufficiently high for each of the corresponding opals such that the PBG is preserved, each additional layer of spheres enhances scattering, which reduces the intensity of transmitted light resulting in a reduction of the PBG minimum $y_c$. Data for opals formed at 31 °C and 34 °C are shown in Supplementary material, Fig. S1. The full optical transmission spectra under all investigated conditions are provided in Supplementary material, Fig. S4.

The opals examined here can be considered periodic-on-average, based on the SEM imaging and Bragg-Snell analysis performed. The optical spectra of such crystals can be interpreted as the result of four primary mechanisms – Anderson localization, absorption, incoherent diffuse scattering, and Bragg diffraction.(46) The effective refractive index of the opals grown here can be approximated using the Drude model based on a two-material system of polystyrene spheres and air in an FCC geometry, resulting in $n_{eff}$ = 1.46. Anderson localization typically requires a much higher refractive index contrast to cause complete localization and any significant impact on the optical properties are unlikely here.(47) Absorption in the considered spectral range for polystyrene can be considered negligible. Therefore, the remaining considerations are coherent Bragg scattering and incoherent diffuse scattering.

For a system with a low refractive index contrast such as artificial opals, it is the (111) close-packed layers of the fcc lattice which provide the greatest contribution to the observed diffraction in the form of the PBG. While decay in transmission intensity occurs with thickness, attenuation of the corresponding stopband, is expected to be enhanced with thickness due to an increase in diffuse scattering. Simultaneously, an increase in the effect of the Bragg diffraction regime is expected with thickness, acting to enhance the depth of the PBG. In the spectra, the PBG attenuation fluctuates while intensity at the stopband center ($y_c$) decreases exponentially. This shows that while Bragg diffraction occurs in a sufficient vein that the stopband retains its spectral shape, indicating that the crystals possess a low concentration of thickness-induced disordered sites compared to close-packed (111) FCC sites. The trade-off between coherent Bragg diffraction and incoherent diffuse scattering is therefore the defining factor in the optical properties observed. While diffuse scattering is particularly evident near the blue edge of the PBG, where Fabry-Pérot resonances are present, coherent Bragg diffraction occurs such that the PBG retains its spectral shape.

For a perfectly periodic system, the optical spectra are expected to evolve with thickness as an attenuation of the stopband which is formed by coherent Bragg diffraction, while frequencies outside the stopband remain close to the incident value.(48) Here, the spectra evolve such that an increase in the diffuse scattering background due to thickness-induced disorder attenuates transmitted intensity at all frequencies, while the disorder is sufficiently low that the Bragg diffraction regime remains dominant, evident in the integrity of the spectral shape of the PBG at all thicknesses.

**Opal Thickness Control**

Formation of opal photonic crystals where the thickness can be controlled is particularly useful when it is to be used as a template for inverse opal formation and the myriad of applications stemming from those structures. Applications based on the opals themselves also require good control of film formation, especially for rapid crystallization and recent efforts have shown that higher concentrations of sphere solutions are required for fast, high quality film formation during the drying phase.(49) Past works have examined the relationship between colloidal crystal thickness and volume fraction during vertical self-assembly.(50, 51) However, a detailed examination of the maximum thickness which can be achieved by these methods, and the conditions under which these thicknesses can be achieved is needed.

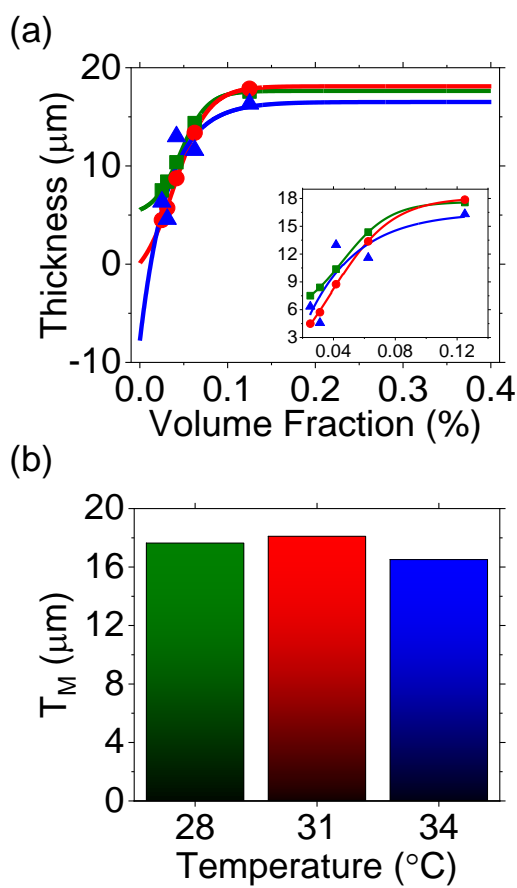

**Figure 2.** (a) Thickness volume fraction relationship for each evaporation temperature with sigmoidal Boltzmann fit function applied to an extended volume fraction range of 0 – 0.4%. Inset: Boltzmann function applied to volume fraction range of 0.025 – 0.125% as per experimental measurements. (b) Maximum thickness $T_M$ for opals grown at each evaporation temperature based on the Boltzmann function.

In Fig. 2 (a), results from the investigations of opal thickness as a function of volume fraction at each evaporation temperature are shown. The values are fitted with the sigmoidal Boltzmann growth function

$$y = \frac{A_1 - A_2}{1 + e^{\frac{x - x_0}{dx}}} + A_2 \tag{4}$$

where $y$ is overall opal thickness, $A_1$ and $A_2$ are opal thickness minimum and maximum values, respectively, $x$ is sphere solution volume fraction, $x_0$ is sphere solution volume fraction at $(A_1 + A_2)/2$, and $dx$ is volume fraction interval constant. The domain of the fit is extended beyond the experimental volume fraction range to 0 – 0.4% to predict the observed thicknesses at very low volume fractions, and maximum thicknesses at high volume fractions. While we observe some scatter in the change in overall thickness at 34 °C owing to the formation of more defective layers in the opal stack at faster evaporation rates, the data indicate a cut-off thickness of ~18 µm for samples grown at each rate, using nominally 370 nm PS spheres over the investigated volume fraction range. At temperatures below 34 °C where the fit is well-behaved, the trend indicates that the colloidal crystal formation occurs sufficiently fast at a solution volume fraction of 0.125% to limit film thickening beyond 37 layers of spheres. Practically, at a volume fraction of 0%, crystal thickness must be zero. For the median evaporation temperature of 31 °C, crystal thickness at 0% from the fit is 0.11 µm and less than the thickness of a single sphere, indicating that the asymmetric sigmoidal function is a reasonably good fit. For the other two temperatures, the fit is far from zero at low volume fractions. In all three cases, the plateau of the fit is very close to the experimentally observed maximum.

For opals grown by EISA, thickness is affected by a myriad of conditions including meniscus shape, substrate tilt angle, relative humidity, and sphere size. However, the two most dominant contributing factors are volume fraction and evaporation rate. For the experiments reported here, the maximum volume fraction of 0.125% was chosen at all three evaporation rates since a breakdown in the self-assembly process was observed for higher volume fractions. In Fig. 2 (b), the maximum thickness ($T_M$) at each evaporation rate based on the sigmoidal Boltzmann fits are shown. The largest thickness is observed at the median temperature of 31 °C, which appears to correspond to the optimal evaporation rate for maximizing thickness while maintaining crystal order at a particular volume fraction. The minimum occurs at the highest temperature of 34 °C. This is the expected result, given that it corresponds to the fastest evaporation rate. For the slowest evaporation rate at a temperature of 28 °C, the sigmoidal fit is good at median to high volume fractions. A deviation from the expected thickness at low volume fractions is expected based on the deviation of the corresponding data from the expected crystal order based on the Bragg-Snell model, which is discussed in the subsequent section. The relatively large magnitude of the $T_M$ values relative to the literature can be attributed to the experimental growth conditions, particularly the employment of the large sphere solution surface areas which allow the use of lower evaporation temperature that result in a more stable assembly process.

**Accuracy of Bragg-Snell Model Under Investigated Conditions**

For highly ordered opals, the reflected wavelengths can be approximated using the Bragg-Snell model. The extent to which each opal obeys the model can be determined using Bragg-Snell plots based on the corresponding angle-resolved transmission spectra. Each plot can be constructed using

$$\lambda_{\text{hkl}}^2 = \frac{4d_{\text{hkl}}^2}{m^2} n_{\text{eff}}^2 - \frac{4d_{\text{hkl}}^2}{m^2} \sin^2 \theta \qquad (5)$$

where $\lambda_{hkl}$ represents the central wavelength of the (111) transmission dip, which constitutes the PBG. The extent to which each plot can be approximated by a linear fit indicates the accuracy of the model.

Figure 3 (a) depicts the agreement between measured average sphere diameter $D_{\text{meas}}$ and the sphere diameter calculated using Eq. 5 for opals grown under all five volume fractions at each evaporation temperature. The difference for each set of conditions is plotted as a function of the adjusted $R^2$ value of the associated Bragg-Snell plots. From examination of the figure, the data can be split into two distinct groups. The first are grouped closely together, with a minimum adjusted $R^2$ value of 0.98 and maximum difference $|D_{\text{meas}} - D_{\text{BS}}|$ of 52 nm, representing a relative difference based on the average sphere diameter for opals grown under all investigated conditions $\Delta D = |D_{\text{meas}} - D_{\text{BS}}|/D_{\text{avg}}$ of ~13 %. Measured values are in close agreement with those calculated, demonstrating the accuracy of Eq. 5 to model the structure of all opals grown, and the high level of order they possess.

The second group are the three outliers, which show a significant deviation from the model based on Eq. 5. For these three sets of conditions, the adjusted $R^2$ values range from 0.83 – 0.97, while the minimum $|D_{\text{meas}} - D_{\text{BS}}|$ is 111 nm, representing a relative difference $\Delta D$ of ~28%. Figures 3 (b)-(d) show the Bragg-Snell plots associated with the three outliers, the opals grown at an evaporation temperature of 28 °C, at the three lowest volume fractions, decreasing in order as 0.042%, 0.031%, and 0.025%. A benefit of constructing Bragg-Snell plots is the ability to directly extract the structural parameters of the corresponding crystal such as interplanar spacing and sphere diameter. The fact that the breakdown in the approximation based on Eq. 5 occurs only for these three sets of conditions indicates that the lower evaporation rate limit for which opals of a sufficient order such that (111) fcc symmetry is achieved is reached here at an evaporation temperature of 28 °C and volume fraction of 0.042%.

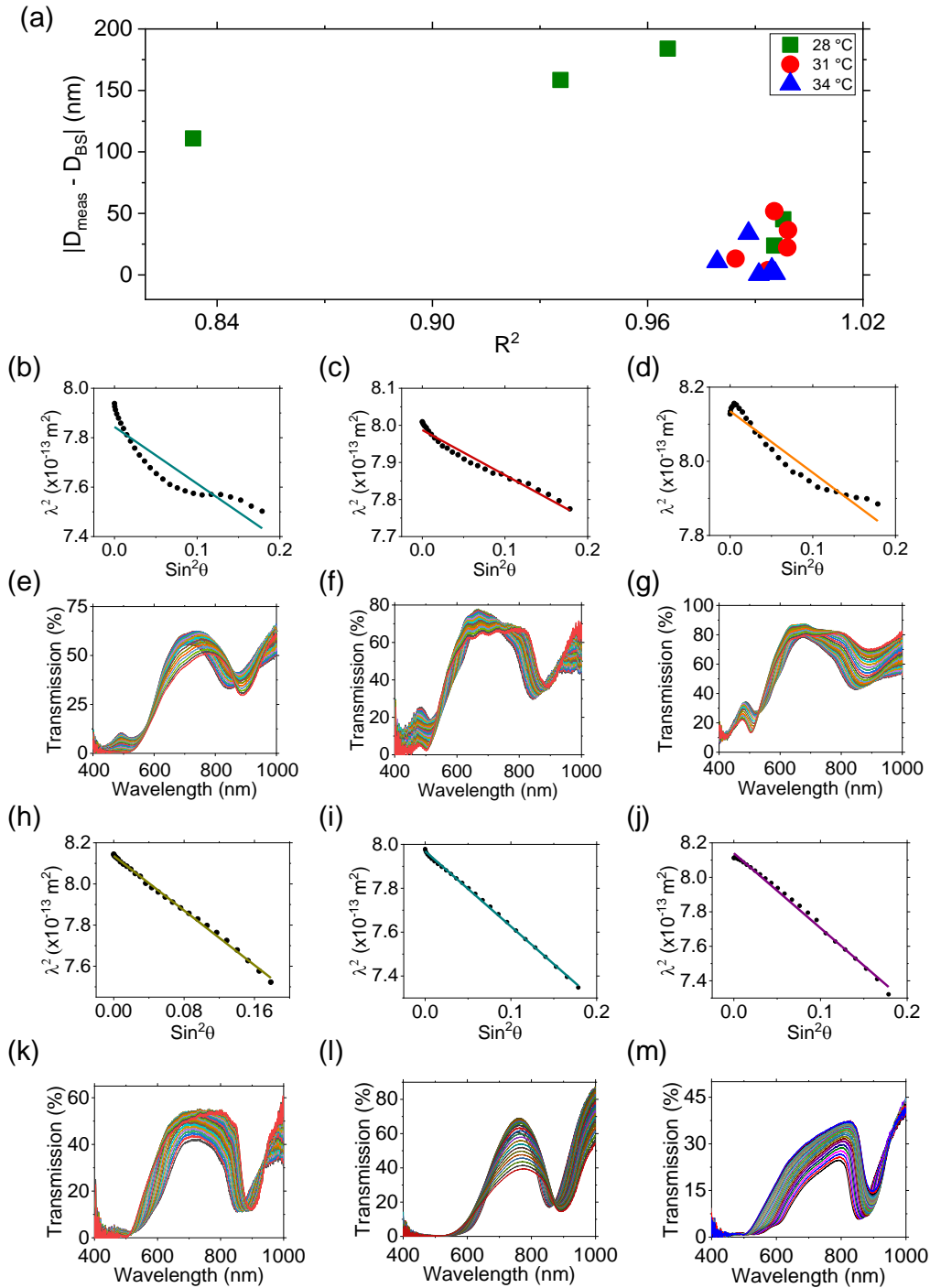

**Figure 3.** (a) Agreement between experimentally measured sphere diameters and those calculated according to Eq. 5 as a function of the $R^2$ value of the corresponding linear fits for opals grown from all five volume fractions at each of the three evaporation rates. (b)-(d) Bragg-Snell plots based on the central frequencies of the primary transmission dips observed in the transmission spectra for opals grown by EISA at an evaporation temperature of 28 °C from sphere solution volume fractions of 0.063% 0.042%, and 0.025% respectively, the three outlier cases for which the Bragg-Snell relationship breakdown occurs. (e)-(g) Corresponding transmission spectra obtained in the incident angle range of 0 – 25° relative to the normal to the (111) plane of an fcc lattice. Bragg-Snell plots for opals grown at evaporation temperatures and volume fractions of (h) 28 °C, 0.031%, (i) 31 °C, 0.042%, (j) 34 °C, 0.025% for three example cases which demonstrate strong agreement with the Bragg-Snell relationship. (k)-(m). Corresponding transmission spectra for the incident angle range of 0 – 25° relative to the normal to the (111) plane of an fcc lattice.

In Figs 3 (e)-(g), optical transmission spectra are shown for an incident angle range of 0 – 25°. The primary transmission dip corresponding to the PBG is observed in all cases. From examination of the spectra, we see the order of these opals deviates from the expected structural order. We observe the appearance of higher order transmission dips from the (220) planes at ~500 nm for all three cases. The most common forms of disorder in the self-assembly of opals are stacking faults.(52-55) While Eq. 5 can be used to calculate the structural parameters of opals based on the (111) transmission dip, it can be used in tandem with Eq. 6, provided in the Supplementary material, to predict structural parameters based on lattice planes of alternative plane orientations. Based on these optical spectra, it appears that below the lower evaporation rate limit for these three lowest volume fractions, breakdown in characteristic (111) fcc symmetry begins and the opal is better able to transmit blue light, after suppression of transmission from the photonic bands associated with the (220) planes in these samples.

In the high-energy region of the spectrum ($a/\lambda > 1$, where $a$ = 563 nm here based on an average measured sphere diameter $D$ = 398 nm), multiple Bragg diffraction by several families takes place and the associated Bragg scattering processes are enhanced. While diffraction from the (220) plane dominates the high energy region, the increase in transmitted blue light is symptomatic of the presence of additional stacking faults.(56) The increase is most pronounced at high incidence angles, which is unsurprising given that stacking faults in the (111) direction have little effect on the optical properties of the crystal, and it is stacking faults in alternative directions which are detected by transmission spectra obtained at angles other than normal incidence.

Another noteworthy aspect of these spectra is that while the PBG is present, there is a clear attenuation of the depth of spectral transmission dips compared to spectra from all other conditions, and a general widening of the PBG occurs. The amplitude of radiation incident on a photonic crystal attenuates rapidly with propagation distance such that the FWHM of the stopband in a colloidal crystal experiences an exponential decay with thickness below the critical thickness, which can be derived using the scalar wave approximation (SWA) and is found elsewhere to be 13 layers for a silica/air colloidal crystal with silica sphere diameters of ~342 nm,(57-59) close in size to the PS opals in this work with an experimentally measured average diameter in the range of 384 – 412 nm (see Figs S5 and S6). For colloidal crystals ranging from single layer to the critical thickness, a narrowing of the stopband occurs similar to the Debye-Scherrer effect in small crystallites and the narrowing of the stopband is inversely proportional to the number of lattice

planes.(60) However, past the point of critical thickness, the narrowing of the stopband saturates. Broadening of the stopband occurs due to disorder in various forms including inhomogeneous thickness and domains corresponding to alternative lattice planes.(61) For the opals analysed here, thickness ranges from a minimum of 10 layers to a maximum of 37 layers. Even crystals of minimum thickness approach the critical value. The broadening of the FWHM is evident as volume fraction and crystal thickness increase, indicative of the inevitable appearance of disordered sites which occurs in the self-assembly process. Light which propagates in these regions of disorder becomes diffuse from scattering, broadening the width of the corresponding stopband.(52)

For comparison, Bragg-Snell plots and transmission spectra corresponding to three of the well-behaved cases are shown in Fig. 3 (h)-(m). For the Bragg-Snell plots, the quality of the linear fit is evident. The transmission spectra show no evidence of higher energy blue photon transmission increases at energies higher than the transmission dip from (220) planes, indicating better overall order. The PBG is also of greater depth and a more well-defined spectral shape, meaning that the structures are dominated by the characteristic (111) symmetry of opals with a consistent and well-defined refractive index contrast throughout the structure, and by comparison to colloidal opal systems with similar spectra response, there is no obvious Fano resonance or Mie scattering effects(62) that contribute to coloration. Branching of the dispersion and overlapping of a PBG from the (111) planes and the (200) planes has been reported previously, and results in curvature of the Bragg-Snell plots(63). In those cases, a specific reflection from the (200) planes was evident due to the high angle spectra of 45° and higher, and no branching of the dispersion and emergence of alternative reflections were seen at lower angle such as we observe. In a separate study, double peaked PBG were linked to stacking faults, but in our case the second dip is located in a separate part of the overall spectrum, and only observed when the volume fraction/temperature pairing forms less that perfect opal films.

The two factors which primarily govern opal order and thickness are sphere solution volume fraction and evaporation rate. While slower evaporation rates tend to facilitate increased crystal thickness, the corresponding volume fraction must be sufficient to prevent spreading of local sphere arrangements such that stacking faults, dislocations and non-close packed arrays are formed. In areas where such disorder occurs, a breakdown in fcc symmetry means that the corresponding crystal deviates from the traditional Bragg-Snell analysis model based on stacking in the (111) direction, and show enhance blue photon transmission after the higher order band gap from (220) planes. For our experiments, an evaporation temperature of 28 °C and volume fraction of 0.042% stands as the critical minimum of this pair of conditions below which the opals lose their characteristic order. For researchers interested in nanostructured inverse opal battery electrodes (64-66) or structured porous materials for sensing electro-photo-catalysis or supports, or for a range of other applications, control of the quality and thickness of the opal template material has a defining impact on the final inverse opal structure.(67, 68)

**Conclusions**

The criteria required for thickness control of opal photonic crystals while simultaneously maintaining a high level of order have been demonstrated here for a range of volume fractions and evaporation rates grown using EISA for thicknesses up to 37 layers of PS spheres. The effect of order and thickness on their associated optical fingerprints are captured experimentally using angle-resolved transmission spectra. The order of each crystal is quantified by the extent to which each opal can be approximated using the Bragg-Snell model for the (111) reflection from an fcc lattice, which constitutes the omnidirectional PBG. SEM measurements are used to directly compare measured structural parameters to those predicted by the Bragg-Snell model. Over the broad range of conditions investigated, those that result in a breakdown in the stacking of an ideal opal are revealed. Furthermore, the maximum thicknesses that can be achieved under the investigated conditions of volume fraction and temperature/evaporation rate are identified. The investigation also demonstrates the extent to which the photonic band gap and other spectral features can be used to assess the degree of (dis)order, structural parameter extraction, and crystal thickness in opal film growth.

**Acknowledgements**

Support from the Irish Research Council under an Advanced Laureate Award (IRCLA/19/118) is gratefully acknowledged. We also acknowledge support from the European Union's Horizon 2020 research and innovation program under grant agreement No 825114. This work is partly supported by Enterprise Ireland Commercialisation Fund as part of the European Regional Development Fund under contract no. CF-2018-0839-P.

# Supplementary Material for

**Dispersion and Thickness Control in Evaporation-induced Self-Assembly of Opal Photonic Crystals**

Alex Grant[1], Alex Lonergan[1], and Colm O'Dwyer[1,2]*

[1]*School of Chemistry, and Environmental Research Institute, and Tyndall National Institute,*

*University College Cork, Cork, T12 YN60, Ireland*

[2]*Advanced Materials and BioEngineering Research Centre, Trinity College Dublin, Dublin 2, Ireland*

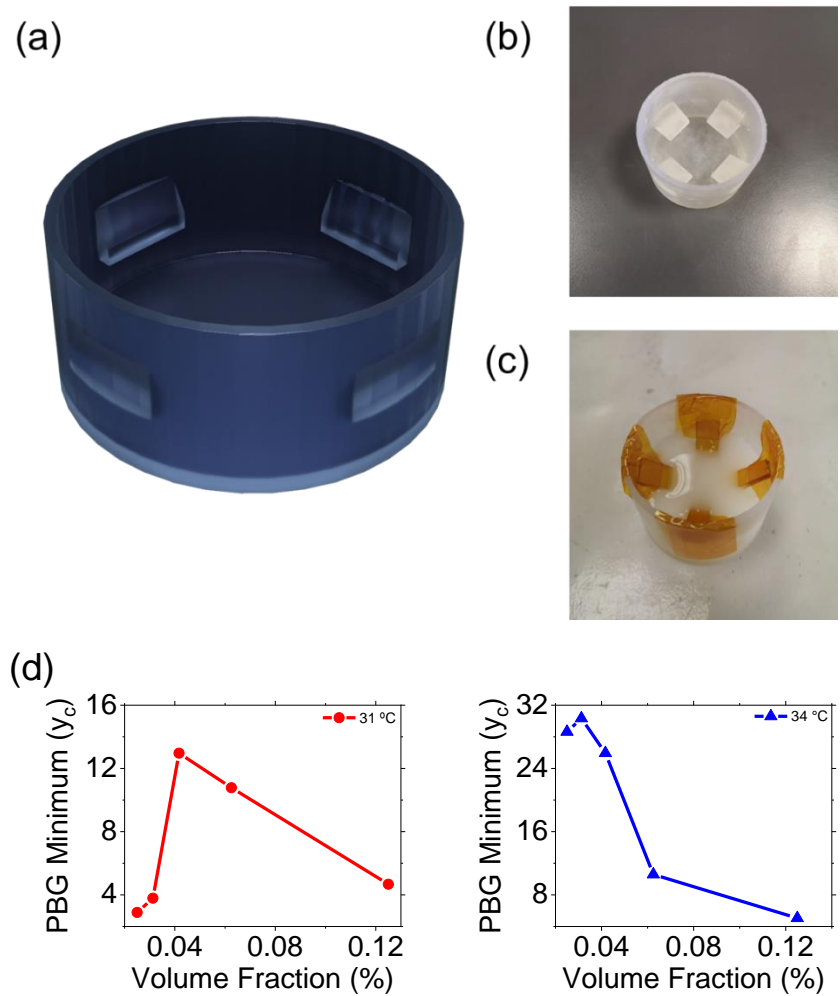

**Figure S1.** SLA 3D printed beaker designed with substrate rests at an angle of 30° to the vertical used for opal growth by EISA. (a) 3D rendered image of beaker. (b) Photograph of beaker. (c) Photograph depicting FTO substrates immersed in PS sphere solution inside beaker. (d) Intensity of central PBG frequency as a function of sphere solution volume fraction for opals grown using EISA at evaporation temperatures of 31 °C and 34 °C.

Fig. S1 shows the 3D printed polymer beakers that were formed from Formlabs Clear Photoreactive Resin using a Formlabs Form 2 stereolithography (SLA) printer. All beakers were cleaned using the same sonication process described above, followed by UV-ozone treatment for 1 h. The diameter of each beaker was 6cm and substrate rests were orientated at an angle of 30° to the beaker walls. Fig. S1 (d) show the relationship between volume fraction and the transmission intensity at the central PBG frequency at normal incidence for opals grown at evaporation temperatures of 31 °C and 34 °C, respectively. A general trend of decreasing PBG minimum occurs as a function of volume fraction for both temperatures. At 31 °C the outliers at low volume fraction significantly affect the trend. For 34 °C, the trend is similar to that observed for the slowest evaporation at 28 °C case.

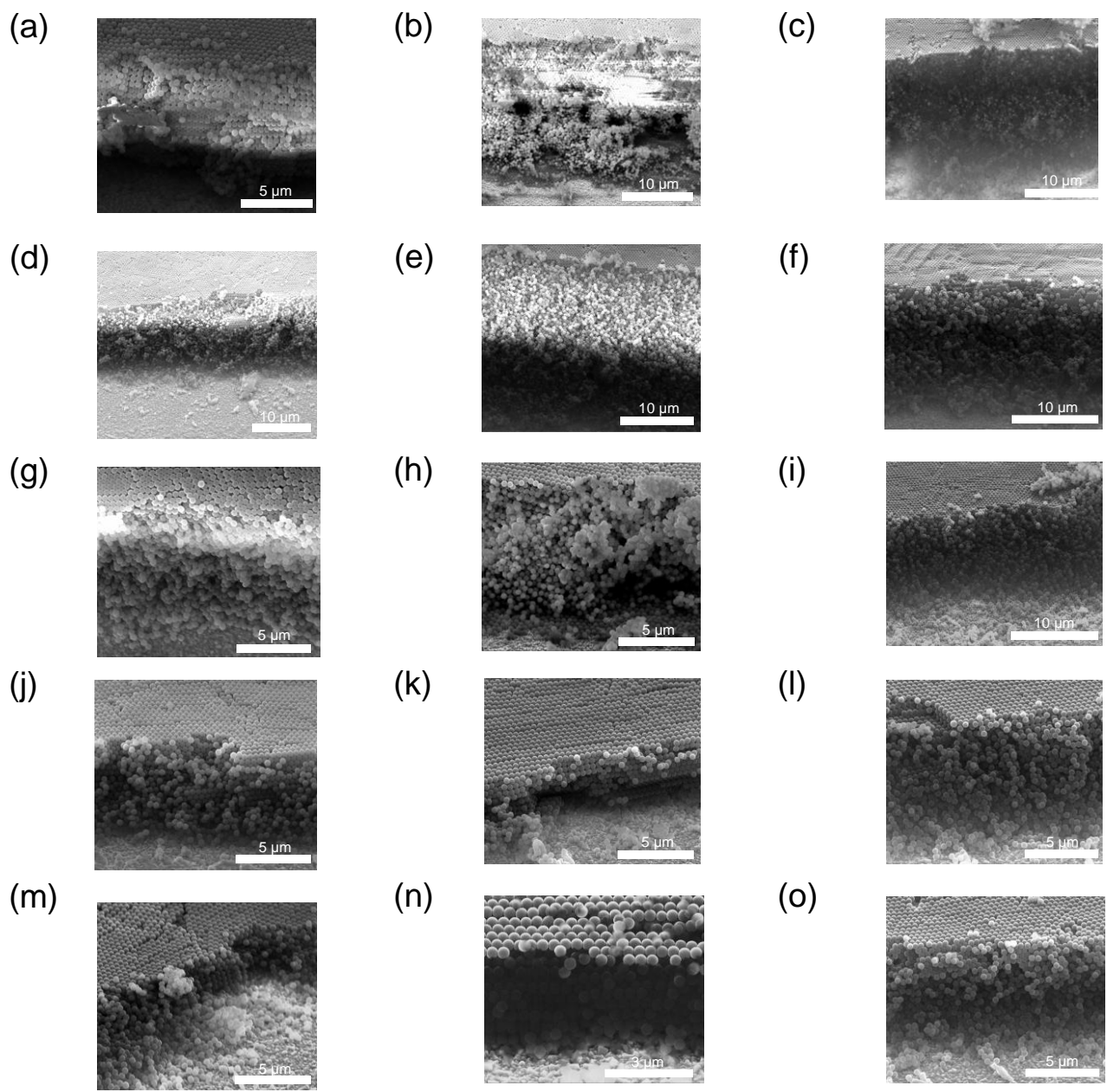

**Figure S2.** Cross-sectional SEM images of opals as examples used for thickness measurements provided to establish relationship with volume fraction. Evaporation temperatures are 28 °C, 31 °C, 34 °C, increasing from columns 1-3. Sphere solution volume fractions: (a)-(c) 0.025%, (d)-(f) 0.0625% (g)-(i) 0.042%, (j)-(l) 0.031% , (m)-(o) 0.025%.

Fig. S2 shows sample SEM images which were used as contributions to calculations of the thickness measurements used as data points in Fig. 3 (a) of the main text. All opals were sputtered with Nickel with a total sputtering thickness of 20 nm prior to imaging to enhance resolution and reduce charging effects.

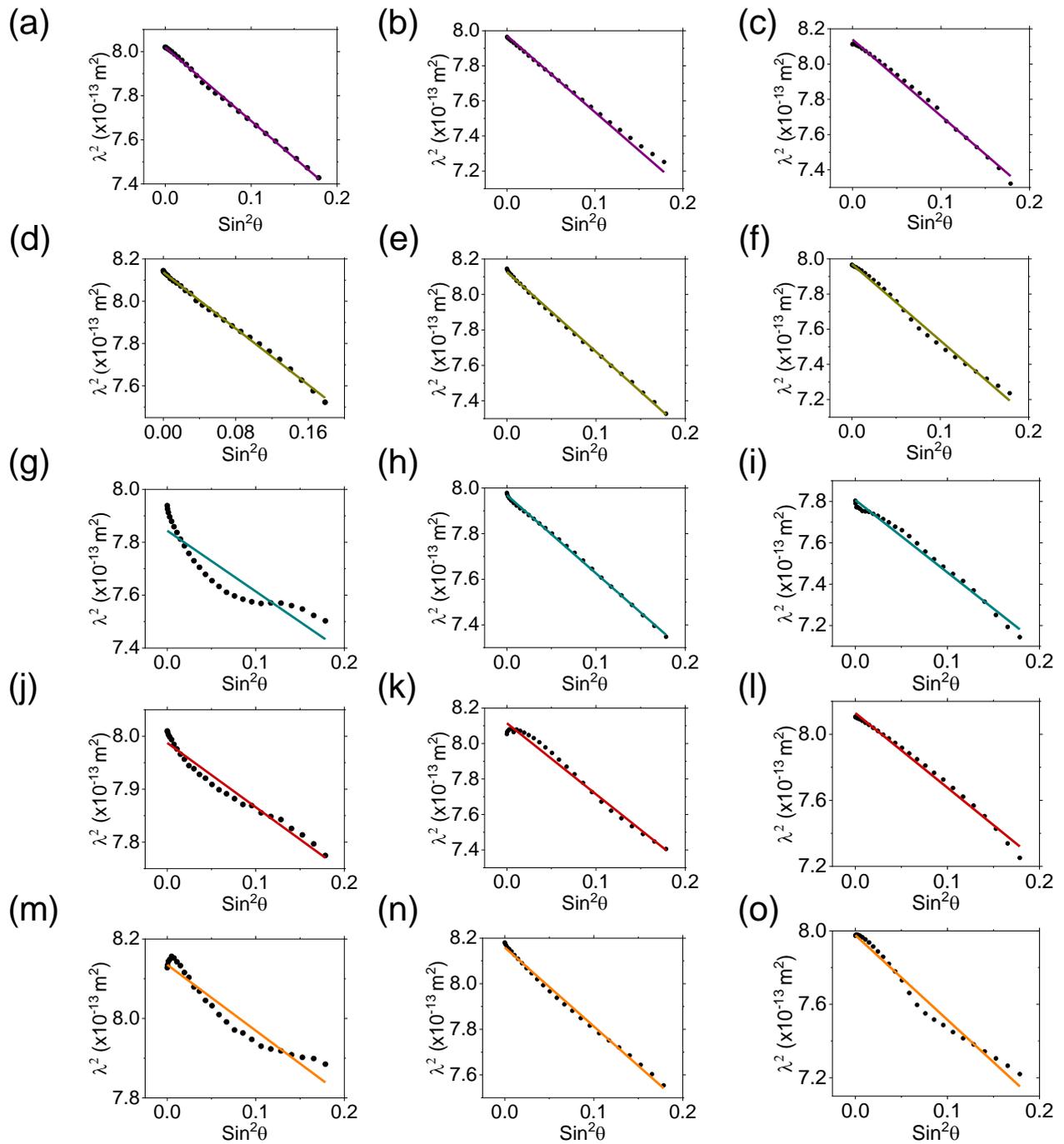

**Figure S3.** Bragg-Snell plots based on the central frequency of the primary transmission dip observed in the optical spectra of opals obtained in the incident angle range of 0-25 °. Evaporation temperatures are 28 °C, 31 °C, 34 °C, increasing from columns 1-3. Sphere solution volume fractions: (a)-(c) 0.125%, (d)-(f) 0.063% (g)-(i) 0.042%, (j)-(l) 0.031%, (m)-(o) 0.025%.

In Fig. S3, Bragg-Snell plots are provided under all investigated evaporation temperatures and volume fractions. Each data point in a Bragg-Snell plot is determined based on the square of the central wavelength of the PBG as a function of the square of the sine of the central wavelength of the PBG is defined as occurring at the minimum of the (111) transmission dip in the associated transmission spectrum. The incident angle is defined relative to the normal to the (111) plane of an fcc lattice. The incident angle range investigated was 0 – 25°. The extent to which each plot can be approximated by a linear fit indicates the dominance of the (111) fcc symmetry and can be used to determine the structural quality of each opal, quantified in this study using the adjusted $R^2$ value.

| Temp (°C) | $D_{meas}$ (nm) | $D_{Eq.\ 3}$ (nm) | $\Delta D/D_{Meas}$ (%) | $d_{Meas}$ (nm) | $d_{Eq.\ 3}$ (nm) | $\Delta d/d_{Meas}$ (%) |
|---|---|---|---|---|---|---|
| 28 | 398.2 | 1046 | 162.68 | 487.69 | 457 | 6.29 |
| 31 | 400 | 439 | 9.75 | 489.9 | 486 | 0.8 |
| 34 | 405.2 | 451 | 11.3 | 496.27 | 493 | 0.66 |

**Table S1.** Comparison of experimentally measurements and calculations using Eq. 3 for opal sphere diameters and interplanar spacings at each evaporation rate.

Table S1 shows a comparison of the structural parameters of sphere diameter and interplanar spacing from experimental measurements and based on calculations using the geometric approach in Eq. 3. Calculated sphere diameter $D_{Eq.\ 3}$ was obtained from the y-intercept of the linear fits applied to each dataset in Fig. 1 (b) of the main text. The calculated interplanar spacing was equated to the slope of each linear fit, where $D_{Eq.\ 3} = \sqrt{\frac{3}{2}} D$. Good agreement between data occurs at the higher two temperatures, while the lowest temperature shows a significant contrast in results.

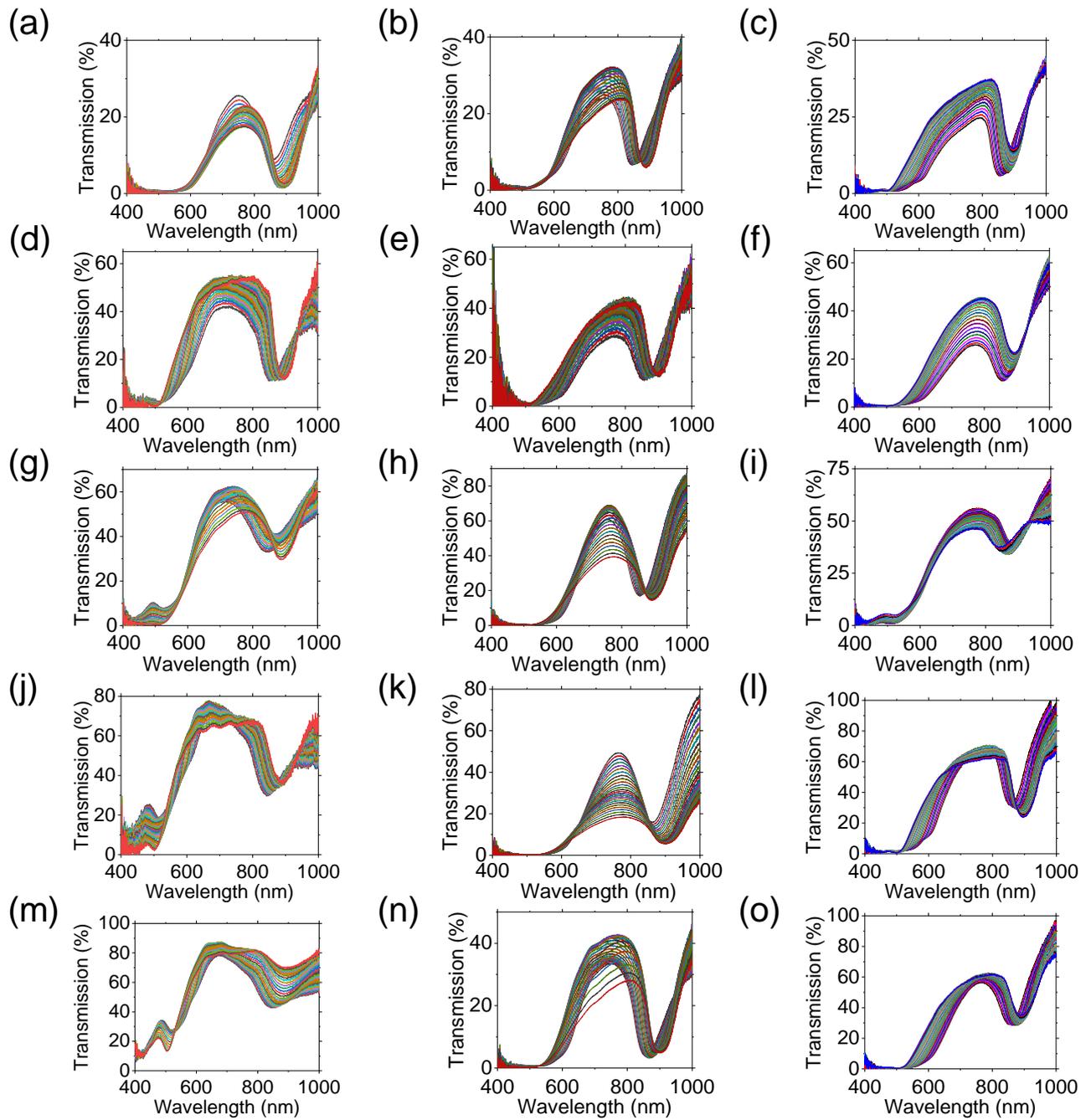

**Figure S4.** Transmission spectra of opals obtained in the incident angle range of 0 – 25° relative to the normal to the (111) plane of an fcc lattice. Evaporation temperatures are 28 °C, 31 °C, 34 °C, increasing from columns 1-3. Sphere solution volume fractions: (a)-(c) 0.125%, (d)-(f) 0.063% (g)-(i) 0.042%, (j)-(l) 0.031%, (m)-(o) 0.025%.

Transmission spectra for opals grown under all investigated volume fractions and evaporation rates in the incident angle range of 0 – 25° are shown in Fig. S2. In all cases, the primary stopband corresponding to the photonic band gap (PBG) is observed. The PBG is blue shifted as incident angle increases. Transmission level at the center of the PBG decreases with increasing volume fraction and hence thickness.

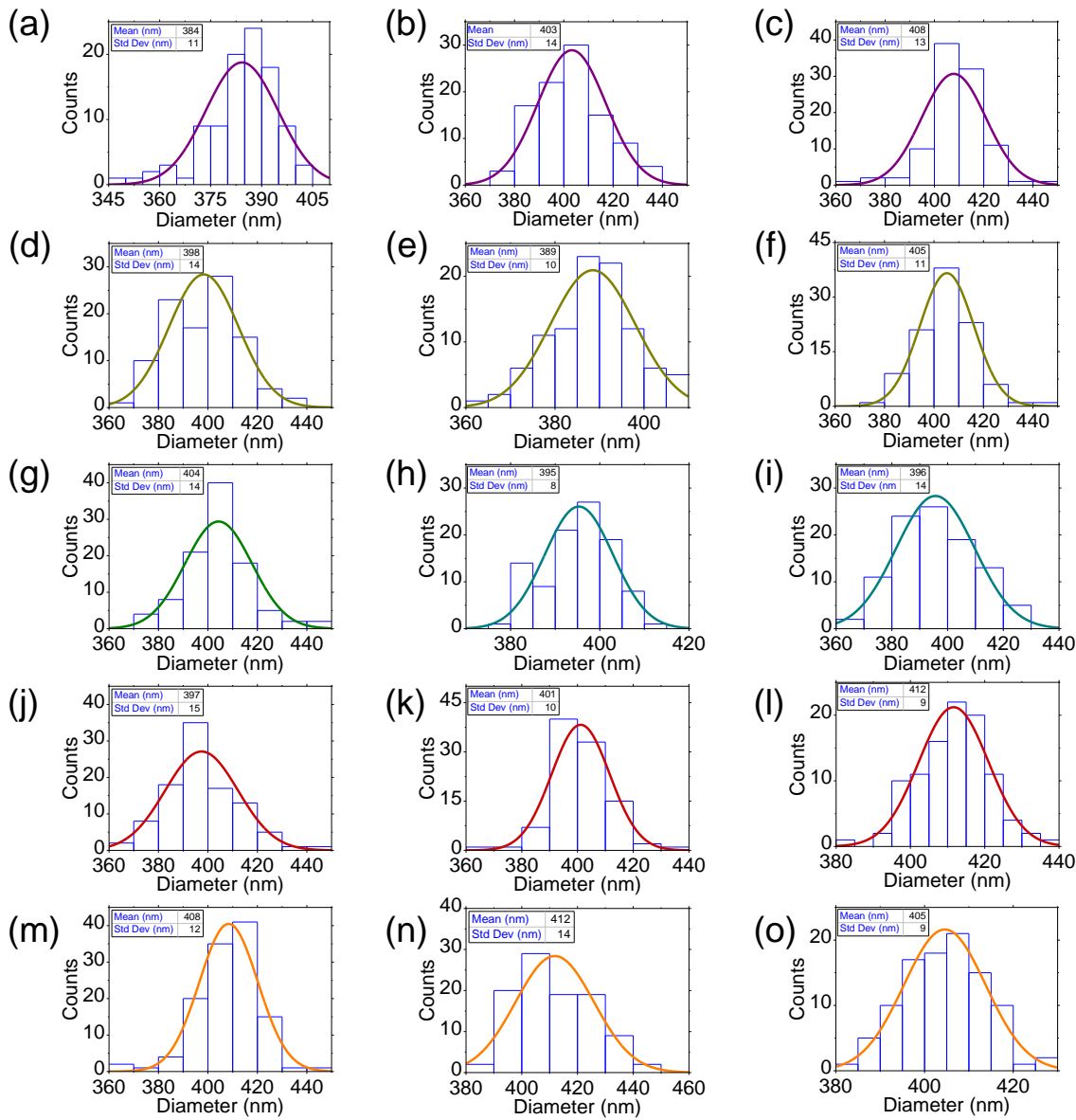

**Figure S5.** Size distributions for opals grown from sphere solutions with a nominal sphere diameter of 370 nm. Evaporation temperatures are 28 °C, 31 °C, 34 °C, increasing from columns 1-3. Sphere solution volume fractions: (a)-(c) 0.125%, (d)-(f) 0.063% (g)-(i) 0.042%, (j)-(l) 0.031%, (m)-(o) 0.025%.

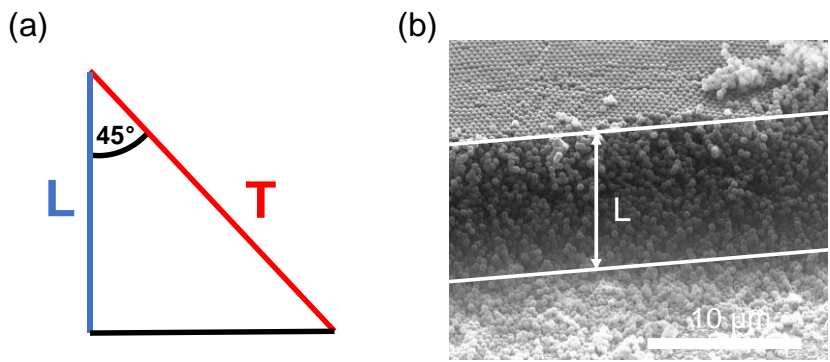

**Figure S6.** (a) Schematic of the determination of opal thickness. (b) Sample cross-sectional SEM image illustrating the method used to acquire thickness measurements.

Thickness measurements were obtained using cross-sectional SEM images. The mathematical method used to determine the actual thickness is shown in Fig. S6 (a), out detailed in the Experimental section of the main text. A sample image is shown as an example in Fig. S6 (b). The white lines depict where the line measurement *L* was determined.

Experimental detection of the PBG, along with any other stopbands which occur in a photonic crystal, can be performed across the full range of the electromagnetic spectrum. Photonic stopbands commonly refer to frequencies from the UV to IR range. However, the iridescent colours characteristic of photonic crystals can only be observed if the stopbands occur in the visible region of the spectrum which is 380 nm – 740 nm.[1] Analysis shows that there is little change in the position of the PBG for opals formed under each set of conditions which can be attributed to their ordered nature and the narrow size distributions. For opals grown here from sphere solutions with a nominal diameter of 370 nm, the PBG lies between 890 nm and 901 nm at normal incidence, corresponding to the near-IR region of the electromagnetic spectrum. A stopband in this range can make no contribution to the observed color in the visible range. Instead, the structural colour of the samples can be explained using Bragg-Snell theory for second order reflections.

While the FCC dominant stacking arrangement causing the (111) transmission dip which forms the PBG is evident throughout this work, secondary Bragg reflections can also be observed in transmission spectra acquired. A sharp increase in transmission can be observed at ~ 485 nm, following higher order transmission dips from the (220) planes. Here, the secondary transmission dips are fully formed only for the two lowest volume fractions with a central wavelength of ~ 500 nm. For the three higher volume fractions, the dips are suppressed. We evaluate the transmission dips for low volume fractions as the result of secondary Bragg reflections due to the most common form of disorder observed in opals, which are stacking faults. We evaluate the disappearance of the transmission dips at higher volume fractions to the attenuation of propagating light in this frequency range as a function of crystal thickness like that experienced by the primary Bragg reflection which constitutes the PBG.

Calculations in the main text have been performed only for reflected frequencies based on growth perpendicular to the (111) plane of an fcc lattice. Secondary Bragg reflections are a result of the interaction of propagating light with alternative lattice planes. Consequently, a correct model for secondary Bragg reflections must incorporate the orientation of the corresponding lattice plane to the (111) lattice plane. The angle α between two such planes defined by the miller indices $(h_1k_1l_1)$ and $(h_2k_2l_2)$ can be calculated using

$$\alpha = \cos^{-1}\left(\frac{h_1 h_2 + k_1 k_2 + l_1 l_2}{\sqrt{h_1^2 + k_1^2 + l_1^2}\sqrt{h_2^2 + k_2^2 + l_2^2}}\right) \qquad (6)$$

The relation above must be incorporated for planes orientated at an angle to the (111) plane. However, the predicted (220) reflection lies parallel to the (111) plane. Artefacts corresponding to allowed Bragg reflections in the (200), (220), (311), and (400) directions may also be observed for self-assembled opals, especially at low incidence angles.[2-4], but are not observed in the spectral range for our samples.